\documentclass[aps,prd,amsmath,preprintnumbers,nofootinbib,a4paper,preprint,floatfix]{revtex4-1}

\usepackage{longtable}
\usepackage{amsthm}
\usepackage{amssymb}
\usepackage{amsfonts}
\usepackage{bbm}
\usepackage{color}
\usepackage{dcolumn}
\usepackage{bm}
\usepackage{pxfonts}
\usepackage{slashed}
\usepackage{graphicx}
\usepackage{multirow}
\usepackage{dsfont}

\usepackage[
  bookmarks]{hyperref}
  
\newcommand{\beq}{\begin{eqnarray}}
\newcommand{\eeq}{\end{eqnarray}}

\newcommand{\bmp}{\noindent\begin{minipage}{16cm}}
\newcommand{\emp}{\end{minipage}\vskip 7mm} 

\def\id{\mathds{1}}

\def\wick#1{\setbox2=\hbox{$\displaystyle#1$}
    \setbox3=\null\ht3=3.0pt\dp3=0.0pt\wd3=20.0pt
    #1\kern-\wd2\kern3.0pt\raise11.0pt\vbox{\hrule height0.3pt
    \hbox{\vrule width0.3pt\box3\vrule width0.3pt}}\kern-24.0pt\kern\wd2}

\def\longwick#1{\setbox2=\hbox{$\displaystyle#1$}
    \setbox3=\null\ht3=3.0pt\dp3=0.0pt\wd3=27.0pt
    #1\kern-\wd2\kern3.0pt\raise11.0pt\vbox{\hrule height0.3pt
    \hbox{\vrule width0.3pt\box3\vrule width0.3pt}}\kern-31.0pt\kern\wd2}

\def\verylongwick#1{\setbox2=\hbox{$\displaystyle#1$}
    \setbox3=\null\ht3=3.0pt\dp3=0.0pt\wd3=43.0pt
    #1\kern-\wd2\kern3.0pt\raise11.0pt\vbox{\hrule height0.3pt

    \hbox{\vrule width0.3pt\box3\vrule width0.3pt}}\kern-47.0pt\kern\wd2}

\definecolor{bluc}{cmyk}{1,1,0,0.1}
\definecolor{rossoCP3}{cmyk}{0,.88,.77,.40}
\definecolor{rosso}{cmyk}{0,1,1,0.4}
\definecolor{rossos}{cmyk}{0,1,1,0.55}
\definecolor{rossoc}{cmyk}{0,1,1,0.2}
\definecolor{verdes}{cmyk}{0.92,0,0.59,0.4}

\hypersetup{colorlinks, bookmarksopen, bookmarksnumbered,
citecolor=verdes, linkcolor=bluc, pdfstartview=FitH, urlcolor=rossos}

\theoremstyle{definition}

\theoremstyle{plain}

\def\lsim{\mathrel{\rlap{\lower4pt\hbox{\hskip1pt$\sim$}}
    \raise1pt\hbox{$<$}}}                
\def\gsim{\mathrel{\rlap{\lower4pt\hbox{\hskip1pt$\sim$}}
    \raise1pt\hbox{$>$}}}                

\newcommand{\ba}{\begin{eqnarray}}
\newcommand{\ea}{\end{eqnarray}}
\newcommand{\be}{\begin{equation}}
\newcommand{\ee}{\end{equation}}
\newcommand{\bd}{\begin{displaymath}}
\newcommand{\ed}{\end{displaymath}}
\newcommand{\een}{\nonumber\end{equation}}
\newcommand{\bea}{\begin{eqnarray}}
\newcommand{\eean}{\nonumber\end{eqnarray}}
\newcommand{\eea}{\end{eqnarray}}

\def\eq#1{Eq.~(\ref{#1})}

\def\fig#1{Fig.~\ref{#1}}
\def\figs#1{Figs.~\ref{#1}}
\def\tab#1{Table~\ref{#1}}

\def\cite#1{\citep{#1}}

\newcommand{\mps}{m_{\rm{PS}}}
\newcommand{\fps}{F_{\rm{PS}}}

\def\mcO{{\mathcal O}}

\def\tr#1{{\mathrm{tr}\left[#1\right]}}

\newcommand{\old}[1]{}

\begin{document}

\title{\Large  \color{rossoCP3} $SU(2)$ Gauge Theory with Two Fundamental Flavours:\\ Scalar and Pseudoscalar Spectrum}
\author{Rudy Arthur$^{\color{rossoCP3}{\varheartsuit}}$}\email{arthur@cp3-origins.net}
\author{Vincent Drach$^{\color{rossoCP3}{\spadesuit}}$}\email{vincent.drach@cern.ch}
\author{Ari Hietanen$^{\color{rossoCP3}{\varheartsuit}}$}\email{hietanen@cp3-origins.net}
\author{Claudio Pica$^{\color{rossoCP3}{\varheartsuit}}$}\email{pica@cp3-origins.net}
\author{Francesco Sannino$^{\color{rossoCP3}{\varheartsuit}}$}\email{sannino@cp3-origins.net}

\affiliation{
\vspace{5mm}
{$^{\color{rossoCP3}{\varheartsuit}}${ \color{rossoCP3}  \rm  CP}$^{\color{rossoCP3} \bf 3}${\color{rossoCP3}\rm-Origins} \& the {\color{rossoCP3} \rm Danish IAS}, University of Southern Denmark, Campusvej 55, DK-5230 Odense M, Denmark \\}
{$^{\color{rossoCP3}{\spadesuit}}$ CERN, Physics Department, 1211 Geneva 23, Switzerland}
 }
\begin{abstract}
We investigate the scalar and pseudoscalar spectrum of the $SU(2)$
gauge theory with $N_f=2$ flavours of fermions in the fundamental
representation using non perturbative lattice simulations. 
We provide first benchmark estimates of the mass of the lightest $0(0^{+})$  ($\sigma$),
$0(0^{-})$ ($\eta'$) and $1(0^+)$ ($a_0$) states, including estimates of the relevant disconnected contributions.
We find $m_{a_0}/\fps= 16.7(4.9)$, $m_\sigma/\fps=19.2(10.8)$ and $m_{\eta'}/\fps = 12.8(4.7)$. These values for the masses of light scalar states provide crucial information for composite extensions of the Standard Model from  the unified Fundamental Composite Higgs-Technicolor theory \cite{Cacciapaglia:2014uja} to models of composite dark matter.
\end{abstract} 
\preprint{CERN-TH-2016-169}
\preprint{CP3-Origins-2016-035 DNRF90}
\maketitle

\section{Introduction}

Models of composite dynamics are often employed to extend the Standard Model (SM) in order to replace the Higgs sector, to describe dark matter or both.
While most of the phenomenological analyses of composite dynamics such as the ones for the composite Higgs are carried out by means of the effective Lagrangian approach, here we determine long sought spectral quantities from the most minimal ultraviolet template theory  \cite{Cacciapaglia:2014uja}. 
Depending on how the model is embedded into the SM, it interpolates between a  technicolor model \cite{Weinberg:1975gm,Susskind:1978ms} and composite Goldstone boson Higgs  \cite{Kaplan:1983fs,Kaplan:1983sm} one.  The model can be further extended  \cite{Cacciapaglia:2015yra}, without introducing elementary scalars, to generate four-fermion interactions able to give mass to the top quark. Partial compositeness \cite{Kaplan:1991dc} is yet another way to generate masses for SM fermions. Large anomalous dimensions of the composite technibaryons (if stemming from purely fermionic fields)  are  invoked. These are, however, hard to achieve  ~\cite{Pica:2016rmv}. One also needs  further model building to connect composite baryons to SM fermions. Following reference \cite{Sannino:2016sfx} one can bypass these hurdles by introducing besides technifermions also techniscalars. If one insists on more involved constructions, to generate SM fermion masses,  featuring only technifermions  then the techniscalars can be viewed as intermediate composite states. The theory template investigated here is again integral part of a key model investigated in \cite{Sannino:2016sfx}.  

The physical Higgs boson is furthermore a mixture of one Goldstone boson and of the lightest scalar excitation, analogue to the $\sigma$ meson in QCD, of the underlying strongly coupled theory.  
With the currently available constraints, such a model is compatible with experiments~\cite{Arbey:2015exa}.
 
We will now use numerical simulations of lattice gauge theories   to study the non-perturbative dynamics of such strongly interacting models and provide first-principles predictions on the spectrum and low energy constants of the theory. 
The main goal of the present work is to determine, using lattice calculations,  bounds on the lightest (pseudo) scalar excitations of the $SU(2)$ gauge theory with $N_f=2$ flavors of Dirac fermions in the fundamental representation, in isolation from the SM.  

Computing the lightest scalar state is notoriously difficult because of the large disconnected contributions to the relevant two point functions, which are extremely challenging to estimate accurately, and because such a state is expected to be a broad resonance decaying into two Goldstone bosons in the chiral limit. 

It is worth noticing that the properties of the scalar resonance of the strong theory in isolation are not  preserved in the full Beyond SM model, due to the many corrections from interactions with the SM gauge bosons and heavy fermions. The mass of the scalar resonance can, for example, become lighter due to the SM interactions \cite{Foadi:2012bb} and consequently narrower for kinematical reasons. 

The scalar sector of strongly interacting theories have been studied in others gauge theories  \cite{Aoki:2013zsa,Aoki:2014oha,Fodor:2015vwa}, but our preliminary results constitute a primer for the important theory investigated here. 
We also provide results on the mass of the pseudoscalar singlet state, the $\eta'$ meson, which is not a Goldstone boson due to the axial anomaly. 
{Such a state, once coupled to the SM fields, decays into two photons \cite{Molinaro:2015cwg,Molinaro:2016oix}  allowing it to be tested at colliders \cite{Molinaro:2016oix}.}

The theory we consider has previously been studied on the lattice and, in particular, it has been shown that the expected pattern of spontaneous chiral symmetry breaking is realized~\cite{Lewis:2011zb}. An estimate of the  masses of the vector and axial-vector mesons in unit of the pseudoscalar meson decay constant have been obtained in~\cite{Hietanen:2014xca}. The scattering properties of the Goldstone bosons of the theory have also been considered~\cite{Arthur:2014zda}, and the model has also been investigated in the context of possible DM candidates~\cite{Hietanen:2013fya,Drach:2015epq,Hochberg:2014kqa,Hansen:2015yaa}.

The paper is organized as follows. We describe in section \ref{sec:techniques} the techniques used in this work for the extraction of the scalar and pseudoscalar mass spectrum and present in section \ref{sec:results} our numerical results. We finally conclude in section \ref{sec:conclusions}.

\section{Lattice techniques}\label{sec:techniques}

We simulate the $SU(2)$ gauge theory with two Dirac fermions in the fundamental representation discretized using the Wilson action for two mass-degenerate fermions $u$, $d$ and the Wilson plaquette action for the gauge fields.
The numerical simulations have been performed using an improved version of the HiRep code first described in Ref.~\cite{DelDebbio:2008zf}.

We use the scale setting and the determination of renormalization constant obtained in \cite{Arthur:2016dir}.  For convenience, we summarize the subset of ensembles used in this work in \tab{table:sim_params}. Even if the flavour symmetry group is $Sp(4)$, we will use the $SU(2)_V$ terminology and thus use the notion of isospin throughout this work. In this work we will focus on fermionic interpolating field operators.

\begin{table}[t]
  \begin{tabular}{ccc}
    \hline
    $\beta$ & Volume & $a m_0$  \\
    \hline
    \hline
    1.8 & $16^3\times32$~~  & -1.00, -1.089, -1.12, -1.14,
    -0.15 \\
     1.8 & $32^3\times32$~~  & -1.155  -1.557\\
    \hline
    2.0 & $16^3\times32$~~  & -0.85, -0.9, -0.94,
    -0.945  \\
    2.0 & $32^4$ &  -0.947, -0.949, -0.-952,-0.957,-0.958 \\
    \hline
    2.2 & $16^3\times32$  & -0.60, -0.65, -0.68
    -0.70 \\
    2.2 & $32^4$ & -0.72,-0.735, -0.75  \\
    2.2 & $48^4$ & -0.76  \\
    \hline
    2.3 & $32^4$ & -0.575,-0.60,-0.625,-0.65,-0.675, -0.685 \\
    \hline
  \end{tabular}
  \caption{Summary of the bare parameters for the numerical simulations used in this work.\label{table:sim_params}}
\end{table}

\subsection{Two-point functions}
%
We define the following interpolating operators:
\begin{eqnarray}
{\cal O}_{\overline{q}q}^{(\Gamma,\pm)}(x) &=& \overline{u}(x)\Gamma u(x) \pm  \overline{d}(x)\Gamma d(x)  \,, 
\end{eqnarray}
where $\Gamma$ denotes any product of Dirac matrices. 

We extract the meson masses from zero-momentum two-point correlation functions:
\begin{align}
C^{(t_0)}_{\Gamma,\pm}(t)
 & =  -\frac{1}{N_f L^3}\sum_{\vec x,\vec x_0} \left\langle {\cal O}_{\overline{q}q}^{(\Gamma,\pm)\dagger}(t,\vec x\,) {\cal O}_{\overline{q}q}^{(\Gamma,\pm)}(t_0,x_0) \right\rangle.
\end{align}
Denoting the quark propagator $S(x,y)$, the Wick contractions read
\begin{align}
  C^{(t_0)}_{\Gamma,-}(t)& = \frac{1}{L^3} \sum_{\vec x,\vec x_0} \langle \tr{S(x,x_0) \Gamma S(x_0,x)\Gamma}      \rangle\, ,
\end{align}
for the iso-vector channel and
\begin{align}
  C^{(t_0)}_{\Gamma,+}(t)& =   C^{(t_0)}_{\Gamma,-}(t) - \frac{N_f}{L^3} \sum_{\vec x,\vec x_0} \langle \tr{S(x,x) \Gamma}^\ast \tr{S(x_0,x_0)\Gamma}      \rangle \, ,
\end{align}
for the iso-scalar channel. 
Note that the overall sign has been chosen such that $C^{(t_0)}_{\gamma_5,-}(t) > 0$. For convenience, we also define the so-called disconnected contribution  as:
\begin{align}\label{eq:Cdisc}
C^{(t_0)}_{\Gamma,\rm disc}(t) & =C^{(t_0)}_{\Gamma,+}(t) - C^{(t_0)}_{\Gamma,-}(t) = - \frac{N_f}{L^3} \sum_{\vec x,\vec x_0} \langle \tr{S(x,x) \Gamma}^\ast \tr{S(x_0,x_0)\Gamma}      \rangle \, .
\end{align}

Next we define the vacuum subtracted disconnected contribution as:
\begin{align}
C^{(t_0)}_{\Gamma,\rm disc, \rm sub}(t) = C^{(t_0)}_{\Gamma,\rm disc}(t) + \frac{N_f}{ L^3}\sum_{\vec x,\vec x_0} \left\langle {\cal O}_{\overline{q}q}^{(\Gamma,+)}(t,\vec x) \right\rangle^\dagger \left\langle {\cal O}_{\overline{q}q}^{(\Gamma,+)}(t_0,x_0) \right\rangle\, ,
\end{align}
and the time source averaged disconnected contribution as:
\begin{align}
 C^{\rm av.}_{\Gamma,\rm disc}(t)= \frac{1}{T} \sum_{t_0} C^{(t_0)}_{\Gamma,\rm disc,\rm sub}(t+t_0)\, .
\end{align}
The full correlator is defined as: 
\begin{align}
 C^{(t_0,\rm full)}_{\Gamma,+}(t) =  C^{(t_0)}_{\Gamma,-}(t) + C^{\rm av.}_{\Gamma,\rm disc}(t)\, .
\end{align}
Finally we introduce one last function: 
\begin{align}\label{eq:Copt}
 C^{(t_0,\rm opt.)}_{\Gamma,+}(t) =  A_{\Gamma,-}\cosh\left[m_{\Gamma,-} \left(\frac{T}{2} - t\right)\right] + C^{\rm av.}_{\Gamma,\rm disc}(t)\, ,
\end{align}
where  $A_{\Gamma,-}$ and $m_{\Gamma,-}$ are obtained by fitting the connected correlator on a given range. This ``optimized'' connected part of the correlator, can be used to build an improved estimator of the full correlator. This can be help to remove excited states contributions if only the connected contribution  receives a significant contribution from excited states. This idea has been introduced in \cite{Neff:2001zr}, and applied in the context of $\eta/\eta'$ mass determination in \cite{Jansen:2008wv,Michael:2013gka,Ottnad:2015hva}.

 We use  $Z_2\times Z_2$ single time slice stochastic sources~\cite{Boyle:2008rh} to estimate the connected part of meson 2-point correlators.  We describe in section~\ref{subsec:disc} how the disconnected contribution is estimated.


\subsection{Effective masses}
We define an effective mass $m^{(\Gamma,\pm)}_{\rm{eff}}(t)$ as  in~\cite{DelDebbio:2007pz,Bursa:2011ru} by the solution of the implicit equation:
\be\label{eq:meff}
\frac{C^{(t_0)}_{\Gamma,\pm}(t-1)}{C^{(t_0)}_{\Gamma,\pm}(t)} = \frac{e^{-m^{(\Gamma,\pm)}_{\rm{eff}}(t)(T-(t-1))} + e^{-m^{(\Gamma,\pm)}_{\rm{eff}}(t) (t-1)}}{e^{ -m^{(\Gamma,\pm)}_{\rm{eff}}(t)(T-t)} + e^{-m^{(\Gamma,\pm)}_{\rm{eff}}(t) t }}\, ,
\ee
where $T$ is lattice temporal extent. At large euclidean time $t$, $m^{(\Gamma,\pm)}_{\rm{eff}}(t)$ approaches the value of the mass of the lightest state with the same quantum numbers as the operator ${\cal O}_{\overline{q}q}^{(\Gamma),\pm}$. In the follwowing we will use  $ C^{(t_0,\rm full)}_{\Gamma,+}$ in  \eq{eq:meff}. If instead $ C^{(t_0,\rm opt.)}_{\Gamma,+}$ is used we will denote the effective mass $m^{(\Gamma,\rm opt.)}_{\rm{eff}}(t)$.


If $m^{(-)} < m^{(+)}$ and $T$ large enough, it is clear from \eq{eq:Cdisc} that $-\log C^{(t_0)}_{\Gamma,\rm disc}(t)  \underset{t\to\infty}{=} m^{(\Gamma,+)} $ and therefore 
\begin{align}
m^{(\Gamma,\pm)}_{\rm eff,\rm disc}(t) \equiv -\log \frac{C^{\rm av.}_{\Gamma,\rm disc}(t)}{C^{\rm av.}_{\Gamma,\rm disc}(t-1)}  \underset{t\to\infty}{=} m^{(\Gamma,+)}  \, .
\end{align}

In the following we will be interested in the case $\Gamma=\id$ and $\Gamma=\gamma_5$. We will denote $m^{(\id,+)}\equiv m_\sigma$, $m^{(\id,-)}\equiv m_{a_0}$, $m^{(\gamma_5,-)}\equiv \mps$ and $m^{(\gamma_5,+)}\equiv m_{\eta'}$.

\subsection{Estimate of disconnected contributions}\label{subsec:disc}
By introducing a set of stochastic volume sources $\xi(x)_{i=1,\dots,N_{\rm hits}}$ and defining $\phi_i = D^{-1} \xi_i$, it is straightforward to build a stochastic estimator of $L^{(\Gamma)}(t)=\sum_{\vec x}\tr{S(x,x) \Gamma}$ :
\begin{align}
L^{(\Gamma)}_i(t) =  \sum_{\vec x}\tr{\xi^\dagger_i(x) \Gamma \phi_i(x)}.
\end{align}
As $N_{\rm hits}\rightarrow\infty$, we have:
\begin{align}
\frac{1}{N_{\rm hits}}\sum^{N_{\rm hits}}_{i=1}  L^{(\Gamma)}_i(t) = L^{(\Gamma)}(t)  + \mcO\left(\frac{1}{\sqrt{N_{\rm hits}}}\right)\, .
\end{align}

We can then build an unbiased estimator of $C^{(t_0)}_{\Gamma,\rm disc}(t)$ as follows:
\begin{align}
C^{(t_0)}_{\Gamma,\rm disc}(t) & \underset{N_{\rm hits} \to \infty} {=} -\frac{N_f}{(N_{\rm hits}/2)^2L^3} \Bigg\langle \sum^{i=N_{\rm hits}/2}_{i=1}\sum^{j=N_{\rm hits}}_{j=N_{\rm hits}/2+1}L^{(\Gamma)\dagger}_i(t)L^{(\Gamma)}_j(t_0) \Bigg\rangle\, ,
\end{align} 
from which it is straightforward to obtain an unbiased estimator of $C^{(t_0,\rm full)}_{\Gamma,+}(t)$.

The number $N_{\rm hits}$ of random sources, or ``hits'', necessary to reduce stochastic noise at the same level or below the gauge average noise depends on the volume, the fermion mass, the lattice spacing and of the number of configuration used.  
In practice we find that $N_{\rm hits}=64$ works well for all our ensembles.


\section{Results}\label{sec:results}

\subsection{Effective masses}

\begin{figure}[h!] 
\centering
\begin{minipage}{.5\textwidth}
  \centering
 \includegraphics[width=\linewidth]{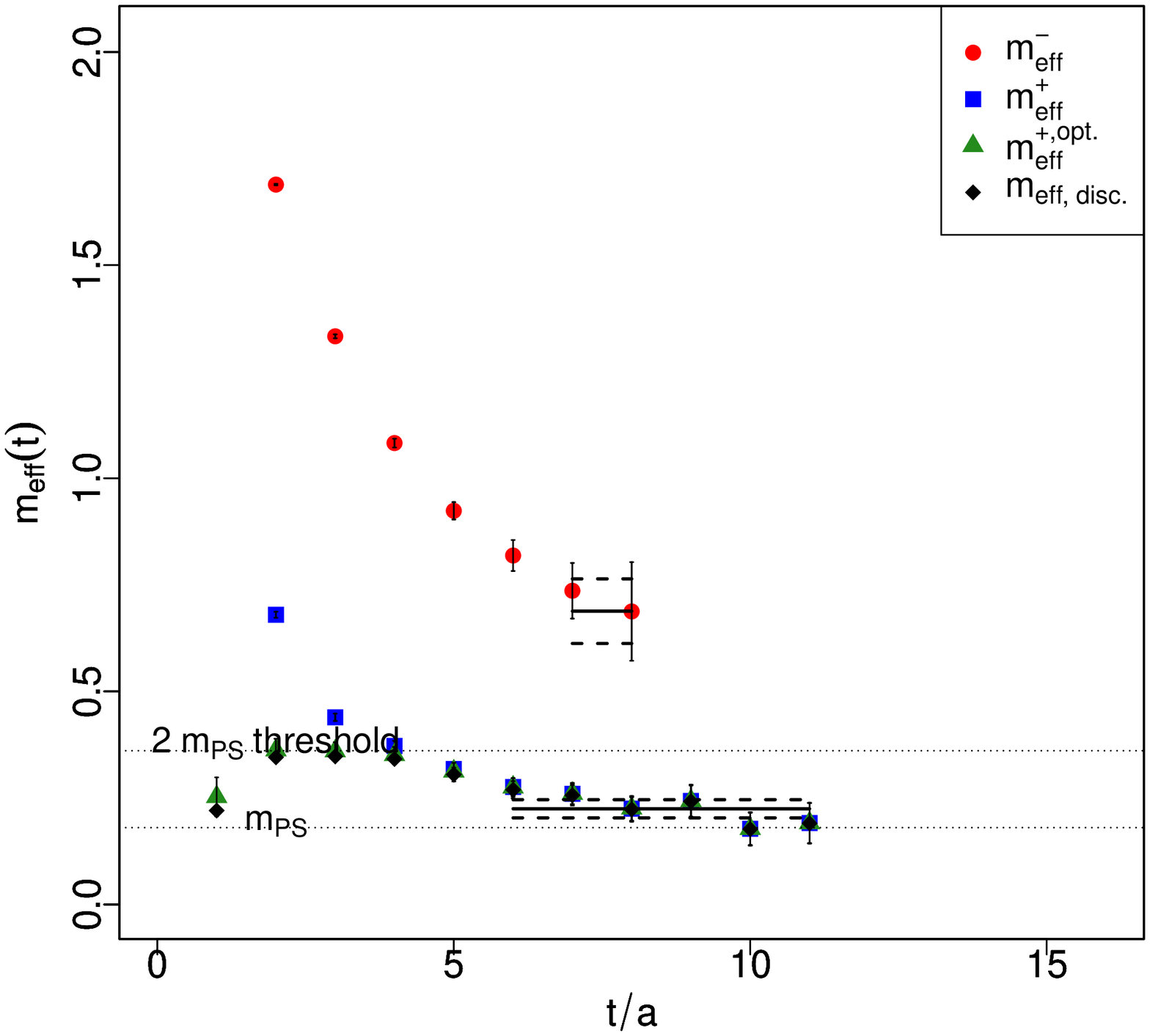}
  \caption{Effective masses of the iso-vector and iso-scalar scalar operator  ($\beta=2.0$, $m_0=0.958$, $L=32$). The disconnected part has been measured on 2200 configurations.}
  \label{fig:meff_s_b20_958}
\end{minipage}%
\hspace*{0.5cm}
\begin{minipage}{.5\textwidth}
\centering
  \includegraphics[width=\linewidth]{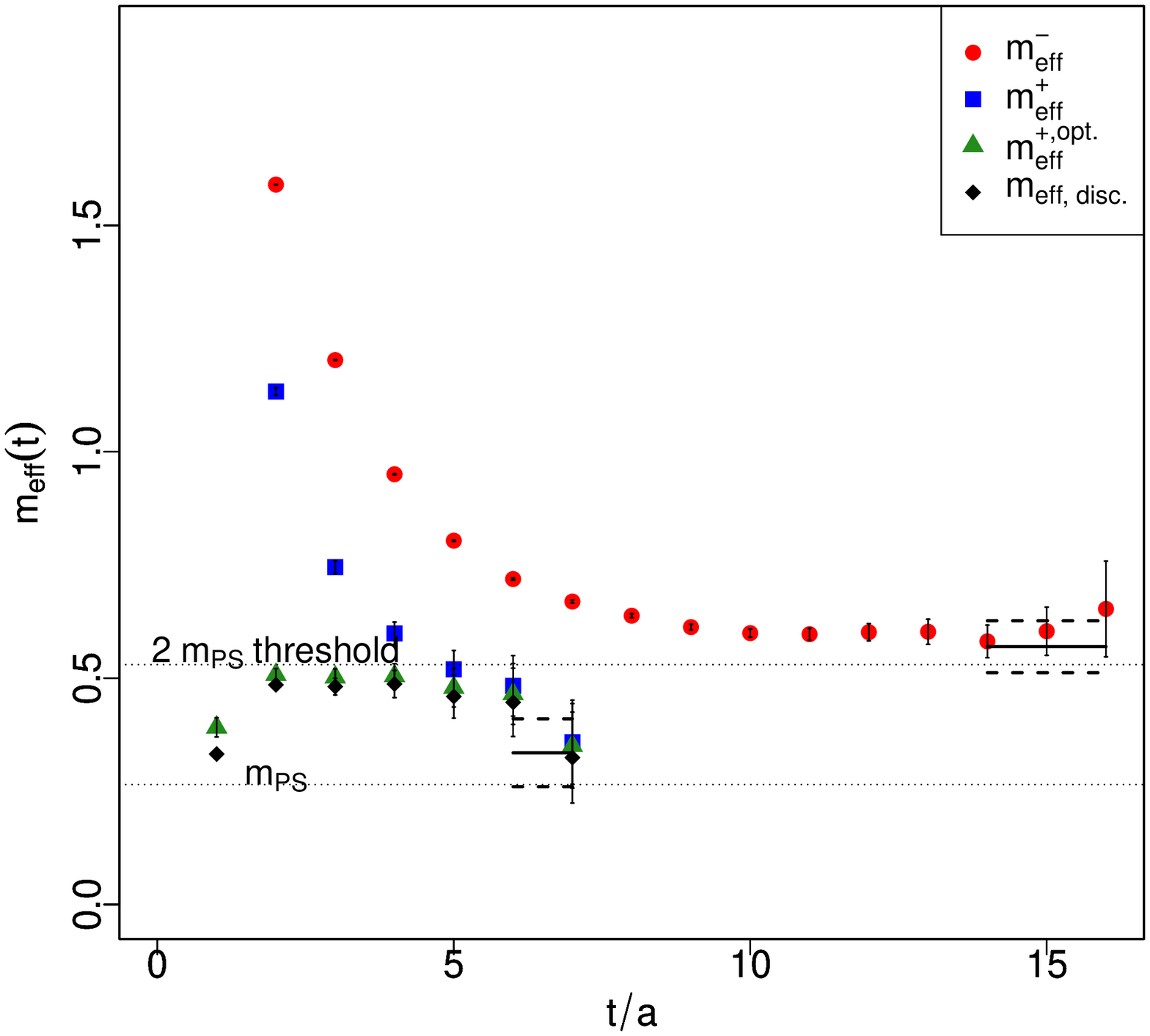}
  \caption{Effective masses of the iso-vector and iso-scalar scalar operator ($\beta=2.2$, $m_0=0.75$, $L=32$). The disconnected part has been measured on 1850 configurations.)}
  \label{fig:meff_s_b22_75}
\end{minipage}
\end{figure}

\begin{figure}[h!] 
\centering
\begin{minipage}{.5\textwidth}
  \centering
 \includegraphics[width=\linewidth]{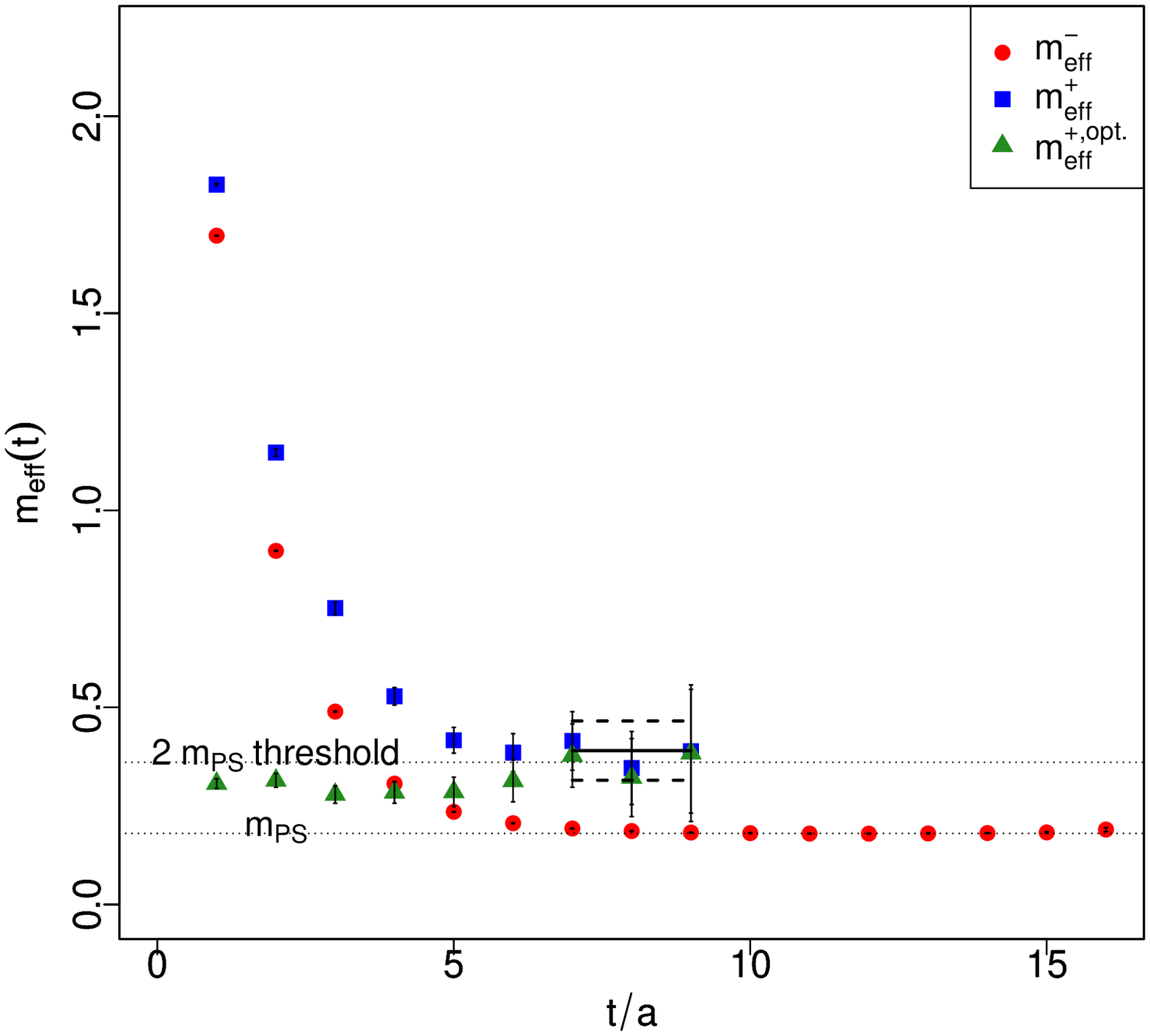}
  \caption{Effective masses of the iso-vector and iso-scalar pseudoscalar operator  ($\beta=2.0$, $m_0=0.958$, $L=32$). The disconnected part has been measured on 2200 configurations.}
  \label{fig:meff_g5_b20_958}
\end{minipage}%
\hspace*{0.5cm}
\begin{minipage}{.5\textwidth}
\centering
  \includegraphics[width=\linewidth]{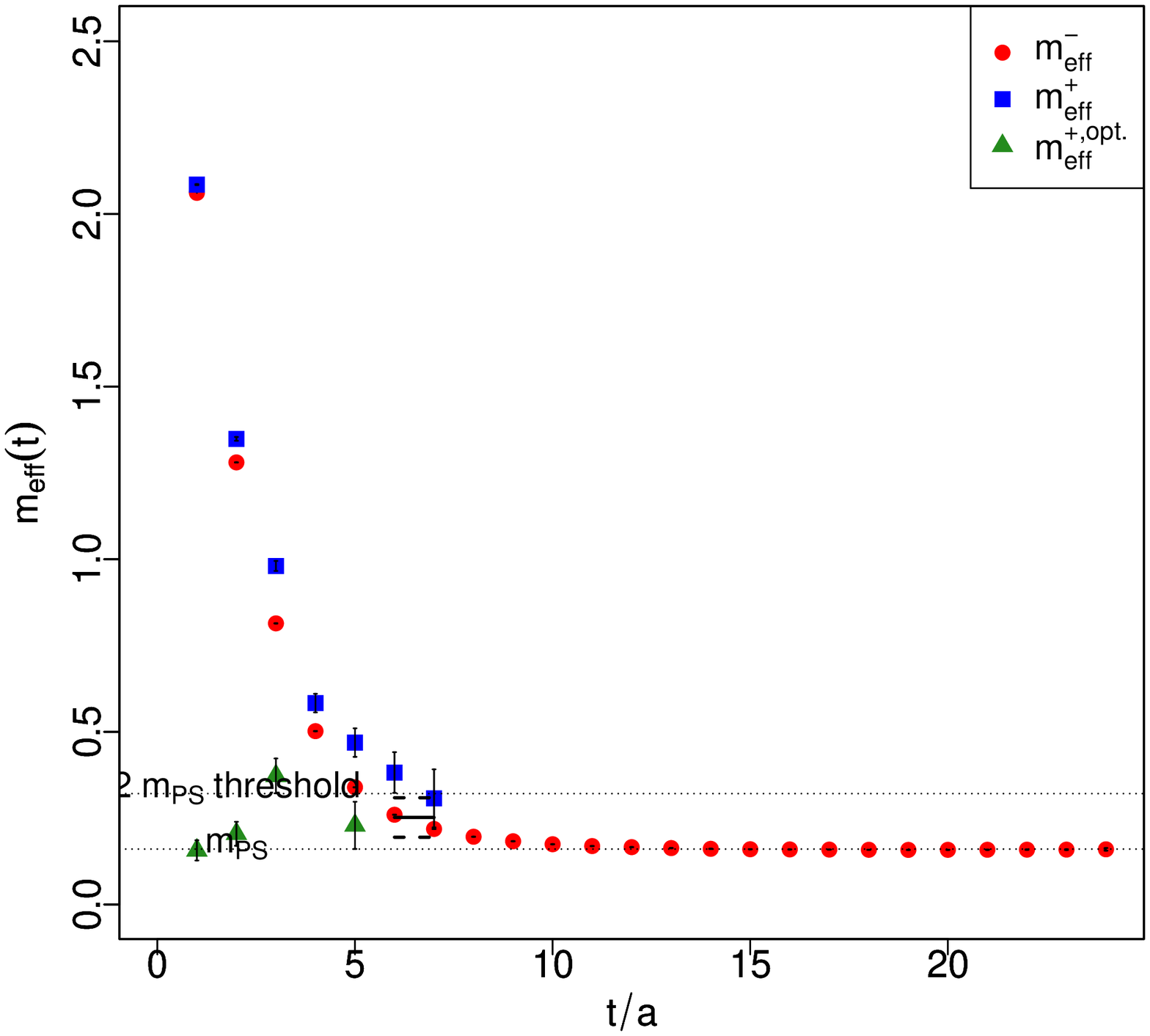}
  \caption{Effective masses of the iso-vector and iso-scalar pseudoscalar operator ($\beta=2.2$, $m_0=0.76$,$L=48$). The disconnected part has been measured on 909 configurations.}
  \label{fig:meff_g5_b22_76}
\end{minipage}
\end{figure}

We illustrate in \figs{fig:meff_s_b20_958},~\ref{fig:meff_s_b22_75},~\ref{fig:meff_g5_b20_958} and~\ref{fig:meff_g5_b22_76} a few examples representative of the quality of the signals we obtain on our most chiral ensembles for two values of the lattice spacing. We plot the effective masses $m^{(\Gamma,-)}_{\rm{eff}}(t)$,  $m^{(\Gamma,+)}_{\rm{eff}}(t)$, $m^{(\Gamma,\rm opt.)}_{\rm{eff}}(t)$ for $\Gamma=\id$ and $\gamma_5$. In the figures, the one and two pions mass thresholds are shown by dotted horizontal lines. 
As for all our ensembles $m_{a_0} > m_\sigma$, we also show $m^{(\id,+)}_{\rm eff,\rm disc}(t)$ in the case of the scalar operator. 

We observe short plateaux before the signal becomes dominated by noise.  
For all our ensembles, the masses are always extracted by performing single exponential fits to the full correlators on a range $[t_1/a,t_2/a]$ where $t_2/a$ is set by the last timeslice where the effective mass is well-defined. We then choose  by inspection the value of $t_1/a$ for each ensemble. 
We show in the figures the mass, its error, and the fitting range obtained from this fit as horizontal lines. 
Note that  $m^{(\Gamma,\rm opt.)}_{\rm{eff}}(t)$ and $m^{(\id,+)}_{\rm eff,\rm disc}(t)$ are only shown for comparison and used as a consistency check.   
The number of thermalized configurations used in each case are reported in the caption of the figures. 

In the scalar channel, we observe that the signal is significantly better for the coarser lattice ($\beta=2.0$) and that the signal over noise ratio improves when $\mps$ is decreased.
Note that in the scalar channel we disregarded the $m_0=-0.76$ ensemble because the signal  is too short with our current statistic. 

In the case of the $\eta'$ meson, we find that at our lighter quark masses the signal clearly depart from the value of the pseudo-Goldstone boson mass $\mps$.


\clearpage
\begin{figure}[h!] 
\centering
\begin{minipage}{.5\textwidth}
  \centering
 \includegraphics[width=\linewidth]{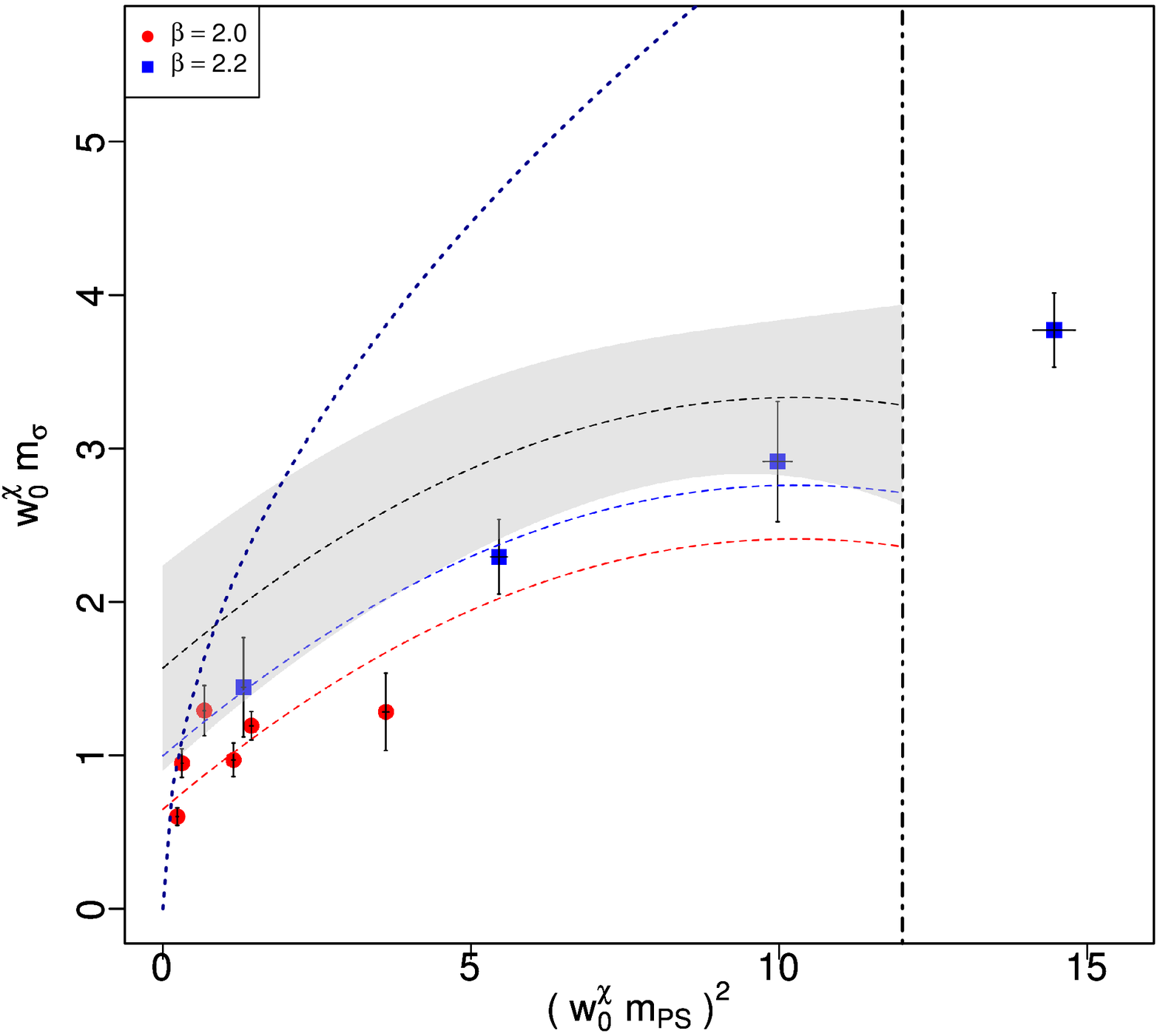}
  \caption{ Combined chiral and continuum extrapolation of the
scalar iso-scalar meson mass $\sigma$. Two pion threshold is depicted by a blue dotted line}
  \label{fig:xfit_sigma}
\end{minipage}%
\hspace*{0.5cm}
\begin{minipage}{.5\textwidth}
  \centering
  \includegraphics[width=\linewidth]{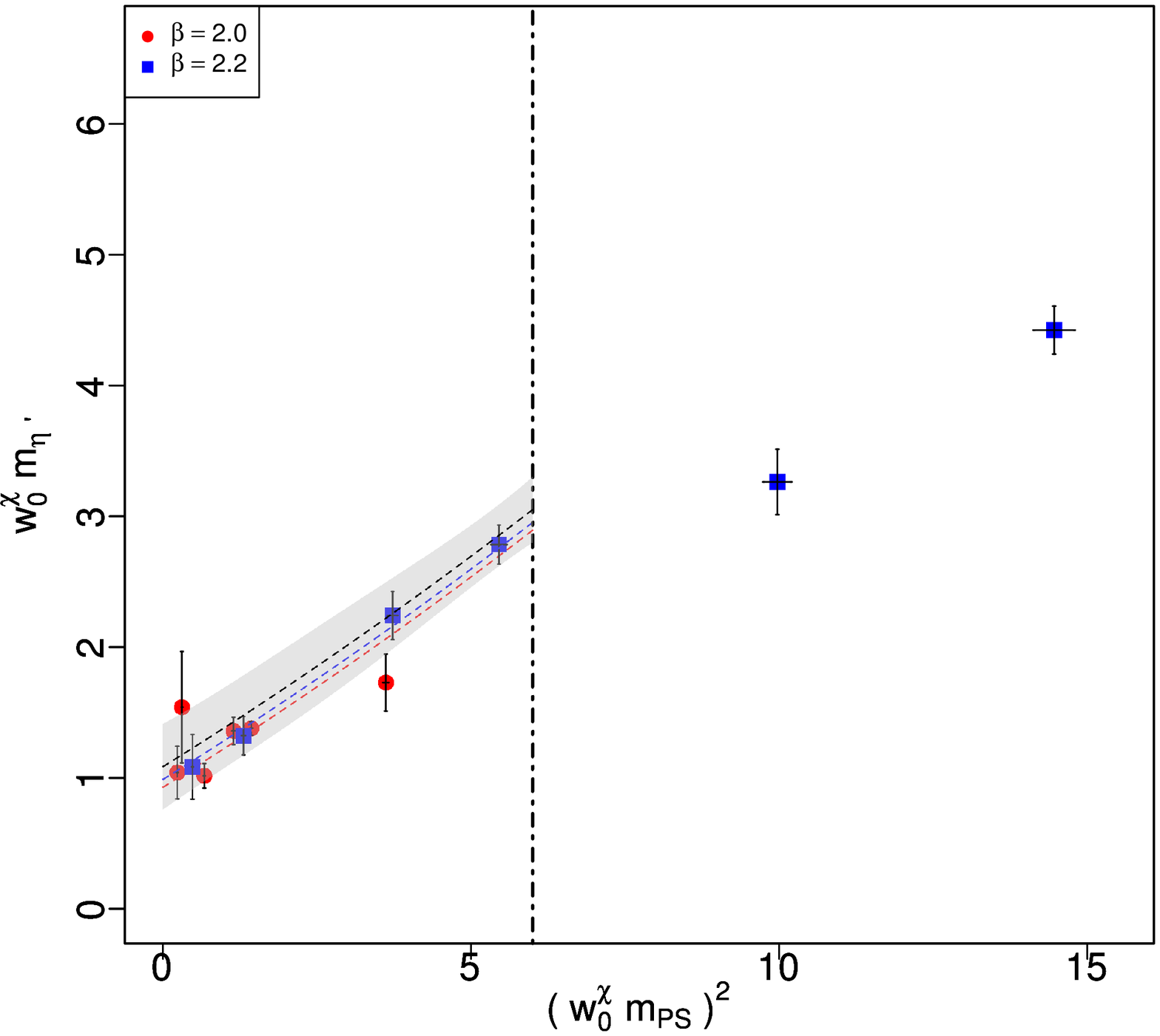}
  \caption{Combined chiral and continuum extrapolation of the
pseudoscalar iso-scalar meson mass $\eta'$.}
  \label{fig:xfit_eta}
\end{minipage}%
\hspace*{0.5cm}
\end{figure}

\subsection{Chiral and continuum extrapolation}
In this section we present the chiral and continuum extrapolation of the $\sigma,\eta'$ and $a_0$ meson masses. All the masses are expressed in units of the scale $w_0^\chi$, as determined in \cite{Arthur:2016dir}, and expressed as functions of $(w^\chi_0 \mps)^2$.  

\begin{figure}[h!] 
\centering
\begin{minipage}{.5\textwidth}
\centering
   \includegraphics[width=\linewidth]{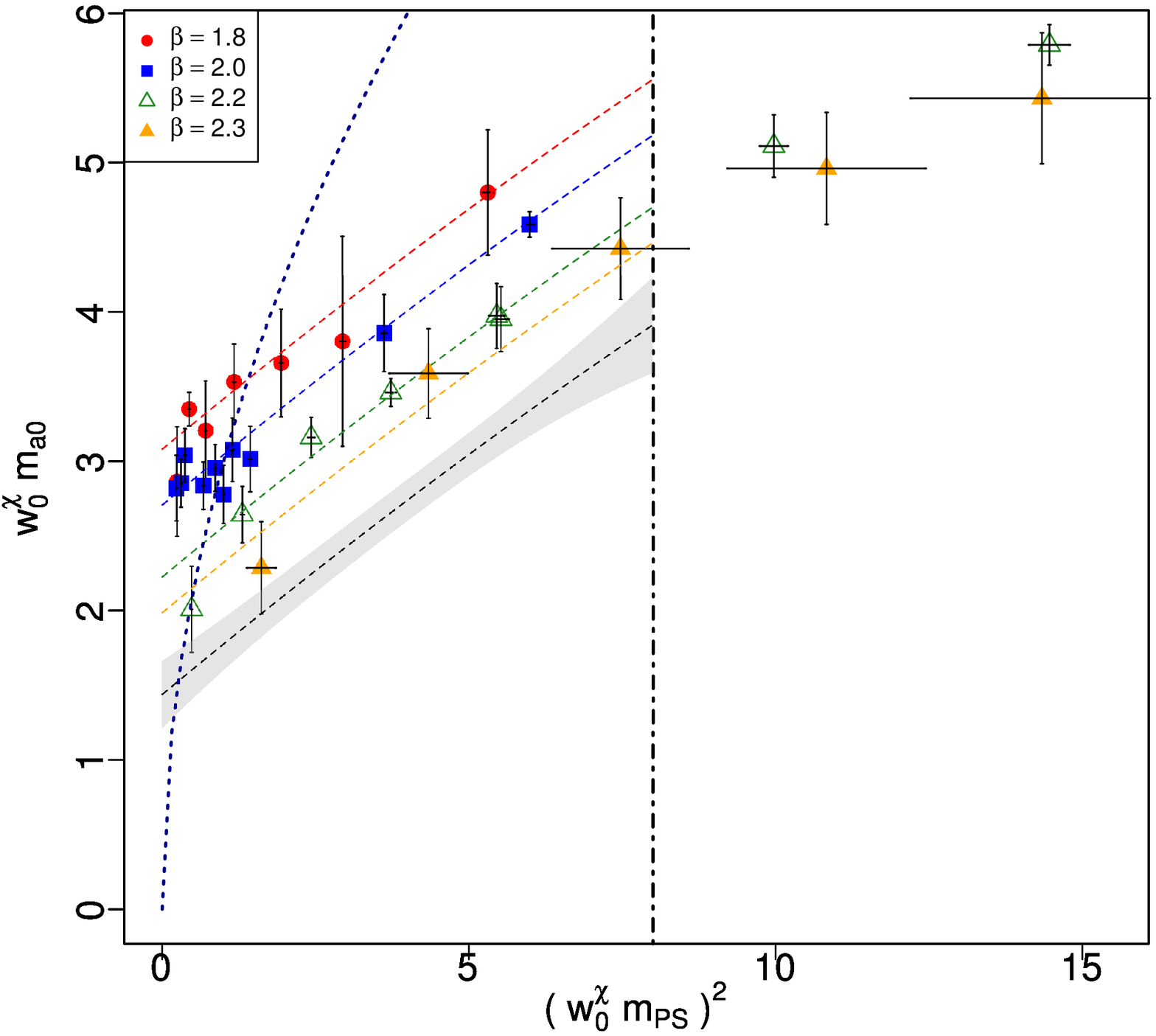}
  \caption{Combined chiral and continuum extrapolation of the scalar iso-vector meson mass $a_0$ using  data at for four lattice spacings. The grey band is our result for the continuum extrapolation and its 1-$\sigma$ confidence region. Data above three  pion threshold are included.}
  \label{fig:xfit_a0}
\end{minipage}%
\hspace*{0.5cm}
\begin{minipage}{.5\textwidth}
\centering
   \includegraphics[width=\linewidth]{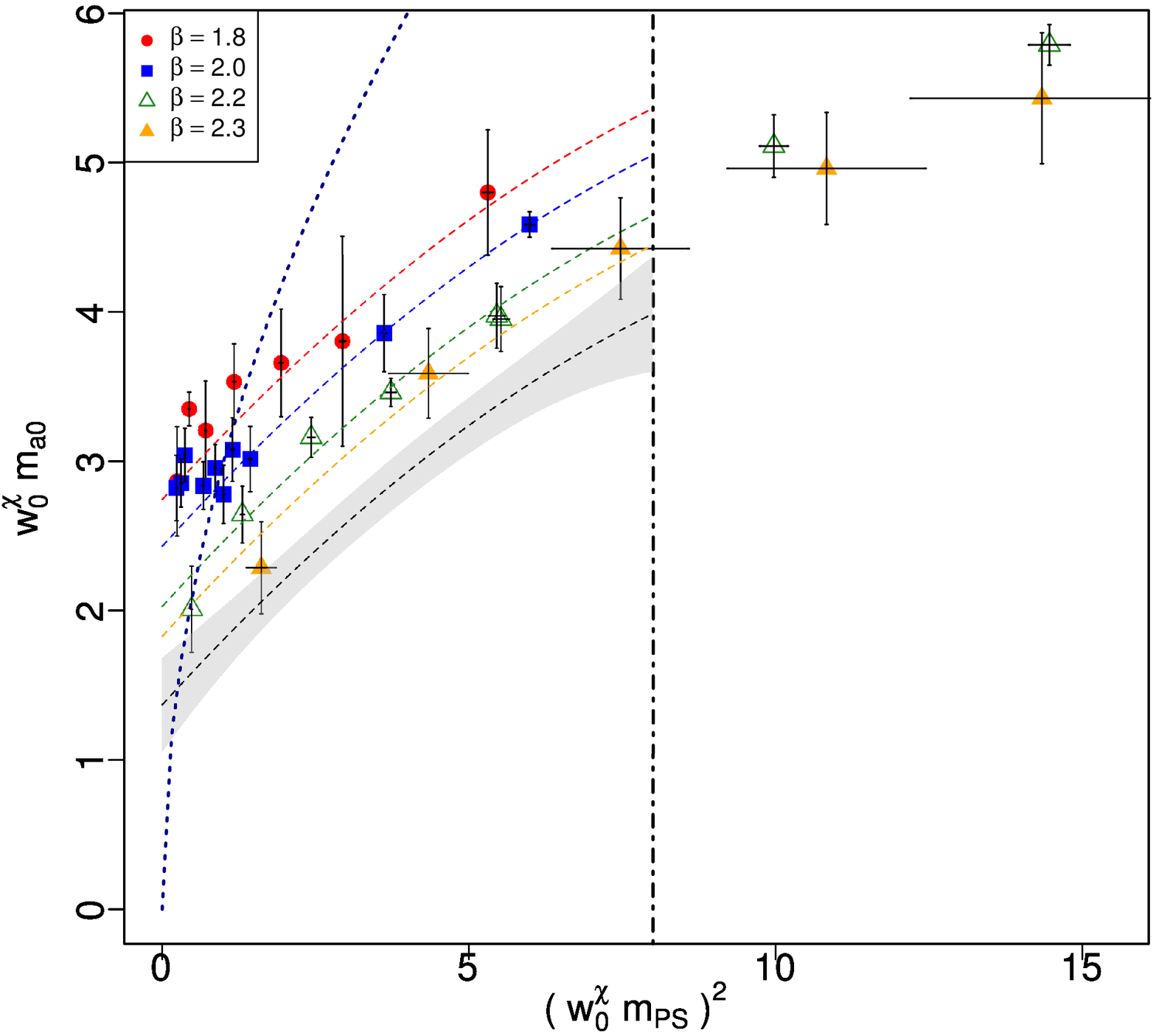}
  \caption{Combined chiral and continuum extrapolation of the scalar iso-vector meson mass $a_0$ using  data at for four lattice spacings. The grey band is our result for the continuum extrapolation and its 1-$\sigma$ confidence region.  Data above three  pion threshold are excluded.}
  \label{fig:xfit_a0_restrict}
\end{minipage}
\end{figure}

The chiral and continuum extrapolation are carried out by using the same strategy as in~\cite{Arthur:2016dir}. For each quantity we perform a global fit, including all the available data, to the following fit ans\"atz:
\be
w_0^\chi m_X = w_0^\chi m_X^\chi + A  (w_0^\chi \mps)^2 + B  (w_0^\chi
\mps)^4 + C \frac{a}{w_0}\, .\label{eq:heavyans}
\ee
The results of the fits for the $\sigma$, $\eta'$ and $a_0$ mesons are shown in \figs{fig:xfit_sigma}, \ref{fig:xfit_eta} and \ref{fig:xfit_a0} respectively and reported in \tab{tab:mX}. In the plots, the gray band indicate the $1\sigma$ confidence region for the continuum prediction, obtained by setting $a=0$ with our best fit parameters.  To give an idea of the fit quality we also plot the best fit curves  at finite lattice spacing using the same color code as for the data points.
The upper limit of the fitting range for each channel is shown by the vertical dashed-dotted line in the plots. 
In the scalar channels, we draw the threshold of the decay into Goldstone bosons by a blue dashed line. In the case of the $\sigma$ meson we thus show that all our results lie below the two Goldstone boson mass threshold. 

In the case of the $a_0$, we have checked that finite volume effects are not significant on three different volumes ($L/a=16,24,32$) at $\beta=2.2$ and $m_0=-0.75$. Since the estimate of the mass of the $a_0$ does not require the estimate of any disconnected loops contribution, we are able to obtain a signal on  all data sets and thus include four lattice spacings in the extrapolation.
However in the case, we observe that some of our data points lie above the 3 Goldstone boson mass threshold.  For this reason, we also consider a fit which excludes the data points above threshold, which is shown in \fig{fig:xfit_a0_restrict}. 

For the $\sigma$ and $\eta'$ channels we also considered constrained fits with fixed $B=0$ in \eq{eq:heavyans}. The corresponding results are also  reported in \tab{tab:mX} and are compatible with the results obtained assuming $B\neq 0$.

We observe that, within our statistical errors, only the $a_0$ suffers from  significant discretization errors.

We take as our final estimates for the scalar meson masses, the results of the fit excluding data above threshold for the $a_0$ channel and the fits of \tab{tab:mX} for the $\sigma$ and $\eta'$. 
We find  $w^{\chi}_0m_{a_0}= 1.3(3)$, $w^{\chi}_0m_\sigma=1.5(6)$ and $w^{\chi}_0 m_{\eta'} = 1.0(3)$. 
By using $w^{\chi}_0\fps=0.078(13)$ from \cite{Arthur:2016dir}, our results can be rewritten in units of $\fps$: $m_{a_0}/\fps= 16.7(4.9)$, $m_\sigma/\fps=19.2(10.8)$ and $m_{\eta'}/\fps = 12.8(4.7)$. 
 
 For comparison in \cite{Arthur:2016dir}, we found $ m_V/\fps=13.1(2.2)$ and  $ m_A/\fps=14.5(3.6)$.

\begin{table}[h!]
  \begin{tabular}{ccccccc}
    \hline\hline
    coef. & $a_0$ & $a_0$  & $\sigma$     & $\eta'$  & $\sigma$     & $\eta'$ \\
    \hline
    type &  all & excluding data above threshold & all  & all & all, $B=0$ & all, $B=0$\\
    \hline
    $w^{\chi}_0 m_X$ &  1.4(2)          &     1.3(3)          &1.5(6)        & 1.0(3)      & 1.4(6)       & 1.0(3)     \\ 
    $A$            &    0.34(8)         &      0.4(1)         &0.34(7)       & 0.2(1)      & 0.22(4)      & 0.32(3)     \\
    $B$            &    -0.00(1)        &      -0.01(1)       &-0.016(8)     & 0.00(2)     & -            & -           \\
    $C$            &    3.3(4)          &      2.8(6)         &-2(1)         & -0.4(8)     & -1(1)        & -0.5(8)     \\
    $\chi^2/$ndof  &    32.06908 / 22   &       13.69283 / 11 &24.97117 / 5  & 12.59433 / 6& 29.69268 / 6 & 13.37 / 7   \\
    cut          &  8          &8     & 12              & 6                      &     12 & 6 \\                  
    \hline\hline
  \end{tabular}
  \caption{Results of the polynomial fits. The cut is in unit of  $(w^{\chi}_0 \mps)^2$ an the upper limit of the fitting range
d. The two last columns show the polynomial fit results assuming $B=0$ \label{tab:mX}}
\end{table}

\begin{figure}[h!] 
\centering
   \includegraphics[width=0.7\textwidth]{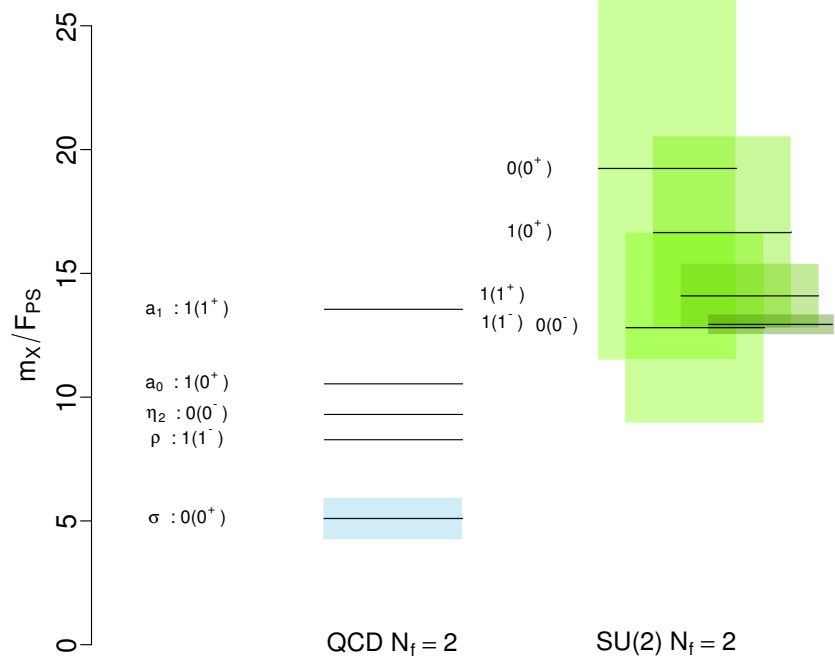}
  \caption{ Comparison of spectrum of QCD ($N_f=2$) with our current
    results. We use the notation $I(J^P)$ to label states. Note that our current results suffer from uncontrolled
    systematic effects which could affect our predictions. Box sizes
    show our error neglecting error on $\fps$. The QCD result are
    taken from experiments (at the physical value of the pion mass)
    except for the $\eta'$ where we took the central value of a
    $N_f=2$ lattice calculation (denoted
    $\eta_2$)\cite{Jansen:2008wv}. The error on the QCD sigma pole mass
    is shown by a light blue band. Note that while in two flavour QCD
    every $I=1$ state is triply degenerate, in our case $I=1$ states correspond to five degenerate states.  }
  \label{fig:spectrum}
\end{figure}

\clearpage
\section{Conclusion}\label{sec:conclusions}

We have presented a determination of the spectrum of the low lying scalar mesons (iso-triplet and iso-singlet) as well as of the $\eta'$ for the SU(2) gauge theory with $N_f=2$ fundamental Dirac fermions.
 
The results are obtained via numerical lattice simulations by using fermionic interpolating operators for the extraction of mass spectrum and include contributions from the disconnected diagrams. 
As expected, the results for the $\sigma$ and $\eta'$ channels receive large contribution from the disconnected part, have an exponentially decreasing signal over noise ratio at large euclidean separations and we observe short plateaux.
Our calculation clearly shows that the $\sigma$ and $a_0$ are stable for most of our ensembles, which provides an {\textit a posteriori} justification of method used to extract the masses of these states in our current setup. 
At lower quark mass, it  will become necessary to consider the two pion scattering process.

 While  we observe large discretization effects by using four lattice spacings for the $a_0$, we do not observe any significant cut off effects for the $\sigma$ and $\eta'$ with two lattice spacings.  
Note that in the range of quark masses explored by our simulations the measured $m_\sigma$ and $m_{\eta'}$ are stable resonances and in most cases this is also the case for the $a_0$.
Assuming that the behaviour of  their mass  as a function of the $\mps$ is not significantly  modified below their respective thresholds, we predict the mass of these states by a polynomial extrapolation to the chiral limit.

We find $w^{\chi}_0m_{a_0}= 1.3(3)$, $w^{\chi}_0m_\sigma=1.5(6)$ and $w^{\chi}_0 m_{\eta'} = 1.0(3)$. 
In units of $\fps$ the results then read:  $m_{a_0}/\fps= 16.7(4.9)$, $m_\sigma/\fps=19.2(10.8)$ and
$m_{\eta'}/\fps = 12.8(4.7)$  using that $w^{\chi}_0\fps=0.078(13)$.  
For comparison we find $ m_V/\fps=13.1(2.2)$ and  $ m_A/\fps=14.5(3.6)$.

For illustrative purposes, we compare in \fig{fig:spectrum}, the low-lying spectrum  for $SU(3)$ and $SU(2)$ gauge theories with two fundamental flavors in unit of $\fps$.  Within our current accuracy, we find that the spectrum is significantly different. 
More numerical simulations are required to better control all systematics and to improve the precision of our findings.

\section*{Acknowledgments}
This work was supported by the Danish National Research Foundation DNRF:90 grant and by a Lundbeck Foundation Fellowship grant. The computing facilities were provided by the Danish Centre for Scientific Computing and the DeIC national HPC center at SDU.
We acknowledge PRACE for awarding us access to resource MareNostrum based in Barcelona, Spain. We thank the Mainz Institute for Theoretical Physics (MITP) for its kind hospitality and support during the meeting \textit{Composite Dynamics: From Lattice to the LHC Run II}, 4-15 April 2016, where part of that work was finalized.

\newpage
\bibliography{paper_scalar}

\end{document}